\documentclass{aa}  

\usepackage{graphicx}
\usepackage{lscape}
\usepackage{txfonts}
\usepackage{xcolor}
\usepackage{lineno}

\usepackage[normalem]{ulem}

\begin{document} 

\newcommand{\mg}[1]{\textcolor{magenta}{#1}}
\newcommand{\barbara}[1]{{\em B: }{\color{red}{#1}}} 

    \title{Constraints on fast radio burst emission in the aftermath of gamma-ray bursts} 


   \author{B. Patricelli
          \inst{1,2,3}
          \and
          M.G. Bernardini
          \inst{4}
          \thanks{The first two authors equally contributed to this work}
          \and M. Ferro
          \inst{4,5}
          }

   \institute{Physics Department, University of Pisa,
              Largo B. Pontecorvo 3, I-56127 Pisa, Italy\\
              \email{barbara.patricelli@pi.infn.it}
         \and
             INFN - Pisa, Largo B. Pontecorvo 3, I-56127 Pisa, Italy
         \and
            INAF - Osservatorio Astronomico di Roma, Via Frascati 33, I-00078 Monte Porzio Catone (Rome), Italy
         \and
            INAF - Osservatorio Astronomico di Brera, via Bianchi 46, I-23807 Merate (LC), Italy\\
            \email{maria.bernardini@inaf.it} 
         \and
            Universit\`a degli Studi dell’Insubria, Dipartimento di Scienza e Alta Tecnologia, Via Valleggio 11, I-22100 Como, Italy
 }        


\date{}

 
  \abstract
   {Fast radio bursts (FRBs) are highly energetic radio transients with a duration of some milliseconds. Their physical origin is still unknown. Many models consider magnetars as possible FRB sources, which is supported by the observational association of FRBs with the galactic magnetar SGR 1935+2154. Magnetars are also thought to be the source of the power of a fraction of gamma-ray bursts (GRBs), which means that the two extreme phenomena might have a common progenitor.}
   {We placed constrains on this hypothesis by searching for  possible associations between GRBs and FRBs with currently available catalogues and by estimating whether an association can be ruled out based on the lack of a coincident detection.}
   {We cross-matched all the Neil Gehrels Swift Observatory (Swift) GRBs detected so far with all the well-localised FRBs reported in the FRBSTATS catalogue, and we looked for FRB-GRB associations considering both spatial and temporal constraints. We also simulated a synthetic population of FRBs associated with Swift GRBs to estimate how likely a joint detection with current and future radio facilities is.}
   {We recovered two low-significance possible associations that were reported before from a match of the catalogues: GRB 110715A/FRB 20171209A and GRB 060502B/FRB 20190309A. However, our study shows that based on the absence of any unambiguous association so far between Swift GRBs and FRBs, we cannot exclude that the two populations are connected because of the characteristics of current GRB and FRB detectors.}
   {Currently available observational data are not sufficient to clearly exclude or  confirm whether GRBs and FRBs are physically associated. In the next decade, the probability of detecting joint GRB-FRB events will be higher with new generations of GRB and FRB detectors, if any: future observations will therefore be key to placing more stringent constraints on the hypothesis that FRBs and GRBs have common progenitors.}


   \maketitle
%

\section{Introduction}

Fast radio bursts (FRBs) are radio bursts with a duration of milliseconds discovered more than a decade ago \citep{2007Sci...318..777L}. To date, hundreds of FRBs have been detected at frequencies ranging between 400 MHz - 8 GHz by several ground-based radio telescopes (see e.g. \citealp{2021ApJS..257...59C}). Some of them have been shown to repeat \citep{2016Natur.531..202S,2019ApJ...887L..30K,2019Natur.566..235C,2023ApJ...947...83C}.
All the observed FRBs have large dispersion measures (DMs), suggesting an extragalactic origin \citep{2007Sci...318..777L,2013Sci...341...53T,2019ARA&A..57..417C}. This has been proven by the identification of the host galaxy of FRB 121102, which is characterised by a redshift of z=0.1932 \citep{2017Natur.541...58C,2017ApJ...834L...7T}. 

The physical origin of FRBs is not established, especially because an observed counterpart at other wavelengths is still lacking (see however \citealt{2016ApJ...832L...1D}). The only exception is the detection of a bright radio burst reported by the \citet{2020Natur.587...54C} and by the Survey for Transient Astronomical Radio Emission 2 (STARE2) radio array \citep{2020Natur.587...59B} that is spatially and temporally coincident with an X-ray burst from the Galactic magnetar SGR 1935+2154 \citep{2020ApJ...898L..29M}. This established the first direct link between magnetars and FRBs. Other evidence supporting the magnetar origin for FRBs comes from the  properties of the host galaxy of the first repeating FRB 121102 \citep{2017ApJ...834L...7T,2017ApJ...843L...8B} and the discovery of a persistent radio synchrotron source that is spatially coincident with it \citep{2017ApJ...834L...8M} and points to a young magnetar born from a catastrophic event, such as superluminous supernovae (SLSNe) and long gamma-ray bursts \citep[GRBs;][]{2017ApJ...841...14M,2017ApJ...843...84N}. This has been confirmed by FRB 20180916B \citep{2020Natur.577..190M,2021ApJ...908L..12T} and FRB 20201124A \citep{2022ApJ...927L...3N}, although FRB 20200120E was localised in a globular cluster \citep{2022Natur.602..585K,2023MNRAS.520.2281N}.

Long and short GRBs have both been suggested to be powered by magnetars \citep{1998A&A...333L..87D,2001ApJ...552L..35Z,2009ApJ...702.1171C,2011MNRAS.413.2031M}. This proposition was shown to be very successful in reproducing different observed properties of the X-ray emission of GRBs, at least from a phenomenological point of view  \citep{2008MNRAS.385.1455M,2011A&A...526A.121D,2012MNRAS.419.1537B,2013MNRAS.430.1061R,2013ApJ...775...67B,2014MNRAS.438..240G,2015JHEAp...7...64B,2018ApJ...869..155S,2023ApJ...949L..32D}. The merger of binary systems of compact objects that contained at least one neutron star were proven to be the progenitors of short GRBs \citep{2017ApJ...848L..13A} and the progenitors of at least a fraction of long GRBs \citep{2022Natur.612..223R}.

Many FRB progenitor models\footnote{For a recent review, see \citet{2019PhR...821....1P} and \url{https://frbtheorycat.org/index.php/Main_Page}.} recently proposed scenarios that indicate a possible association with GRBs. These models can be grouped into two main categories: non-catastrophic models (mainly related to repeating FRBs) and catastrophic models (mainly related to non-repeating FRBs). Non-catastrophic  models include magnetar flares (e.g. \citealp{2014ApJ...797...70K,2018ApJ...868L...4M,2024arXiv240704095D}) and  giant pulses from neutron stars \citep{2016MNRAS.457..232C}, and catastrophic models include the collapse of supramassive neutron stars \citep{2014ApJ...780L..21Z,2014A&A...562A.137F,2016MNRAS.459L..41P} and the merger of compact stars \citep{2016ApJ...826...82L}. Depending on the scenario considered, the FRB could occur from milliseconds before to up to years after the GRBs.

Several searches for a systematic association of FRBs with GRBs have been performed in the last years (e.g. \citealp{2019A&A...631A..62M,2020A&A...637A..69G,2023ApJ...954..154C,Ashkar:2023E8}), and only two possible associations were found so far: FRB 20171209 with the long GRB 110715A, whose afterglow is consistent with being powered by a magnetar \citep{2020ApJ...894L..22W}, and FRB 20190309A with the short GRB 060502B \citep{2024ApJ...961...45L}. The lack of other coincident detections might be due to the sensitivity of high-energy instruments (e.g. \citealt{2020A&A...637A..69G, 2021ApJ...921L...3M}) or to different beaming angles for the radio and high-energy emission (e.g. \citealt{2021MNRAS.501.3184S}). 

In this work, we perform a new systematic search for a GRB-FRB association using the most recent catalogue of FRBs, which collects data obtained with the Canadian Hydrogen Intensity Mapping Experiment (CHIME) instrument, and the sample of all GRBs detected by Swift so far. We used the precise localisation of GRB afterglows in our search, and we allowed a few years of time delay between a GRB and an FRB. The paper is organised as follows: in Sect. \ref{sec:archival} we describe our search method and the results we found; in Sect. \ref{sec:limits} we assess the significance of the results of our search to confirm or exclude the association between GRBs and FRBs; and finally, we discuss in Sect. \ref{sec:conclusions} our results and summarize our conclusions.
 
\section{Search for an association between gamma-ray bursts and fast radio bursts using archival data}\label{sec:archival}

We searched for possible associations between GRBs and FRBs using currently available catalogues.

\subsection{Cross-match between GRB and FRB catalogues}\label{sec:archivalsearch}
We selected\footnote{\url{https://swift.gsfc.nasa.gov/archive/grb_table/}} all the GRBs (both short and long) that were detected by the Neil Gehrels Swift Observatory (Swift) until March 2023 for which there is an X-Ray Telescope (XRT) detection and for which based on this, the position is available with an accuracy $\sigma_{\rm GRB} \lesssim 5''$ (1276 GRBs). We also drew from this sample a subsample  of GRBs for which a reliable redshift measurement was available (400 GRBs). In order to extend the time window of the research, we also considered a sample of pre-Swift GRBs from BeppoSAX, the High Energy Transient Explorer II (HETE-II), and the INTErnational Gamma-Ray Astrophysics Laboratory (INTEGRAL), which have a localisation with an accuracy of $\sigma_{\rm GRB} < 10''$ and a redshift (32 GRBs; the first is GRB 970228).

We considered all the FRBs from the  FRBSTATS Catalogue\footnote{\url{https://ascl.net/2106.028}} \citep{2021ascl.soft06028S} available until March 2023 (828 FRBs; the last is FRB 20221128A), both repeating and not-repeating. From these, we selected the FRBs with an accuracy in the localisation\footnote{Although the error regions of FRBs are elliptical, for the well-localised case, the distribution of $\delta = (\sigma_{RA}-\sigma_{Dec})/\sigma_{Dec}$ has a mean value $\mu_\delta=0.15$ and standard deviation $\sigma_\delta=1.1$. Thus, in what follows, we approximate the error regions as circular.} $\sigma_{\rm FRB} \leq 30'$ (633 FRBs).

We then searched for any Swift GRB that is spatially coincident with an FRB from the two samples described above, without any further restriction. In order to have a coincidence, we required that the distance between the two is smaller than 3$\sigma$, with $\sigma^2=\sigma_{\rm GRB}^2+\sigma_{\rm FRB}^2$, and in any case, not larger than $30'$. We find 28 matches. However, the most likely scenarios are that the FRB is either coincident or follows the GRB event. 

\begin{figure*}
\sidecaption
\includegraphics[width=12cm]{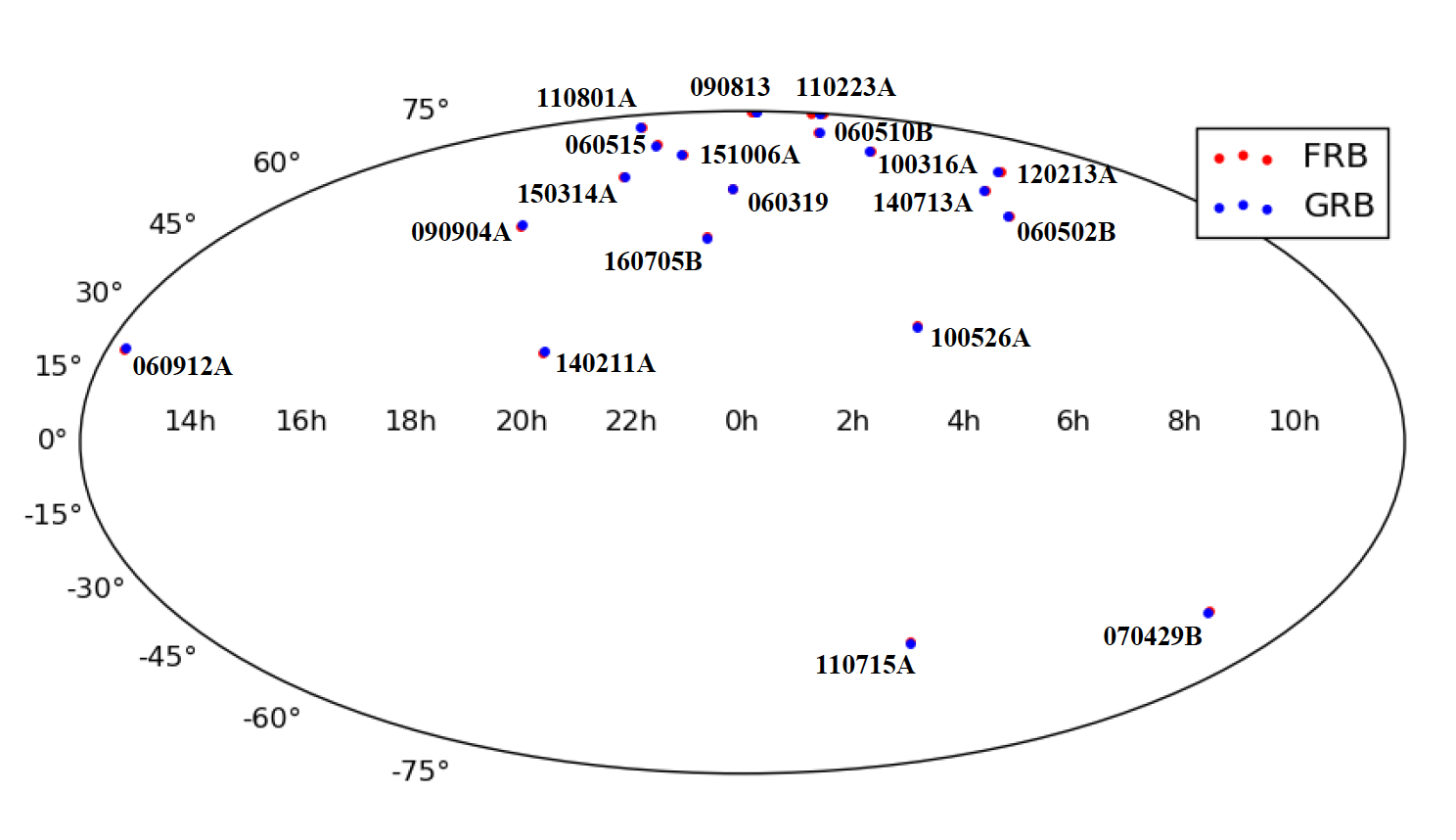}
\caption{FRBs and GRBs with a positive match. The GRB names are also shown.}\label{fig:match}
\end{figure*}

When we concentrate on the cases in which the FRB follows the GRB event, we find 21 the positive matches. These are shown in Fig. \ref{fig:match} and listed in table~\ref{tab:list}. All the positive matches involve non-repeating FRBs. In two cases, the same GRB matches two different close-by FRBs (GRB 110223A with FRB 20190519H and FRB 20190609A, and GRB 090813 with FRB 20190425B and FRB 20190609B).

For GRBs with a redshift (6 out of 21 spatially coincident GRBs with an FRB), we can also use the information on the distance and compare it to the inferred redshift of the corresponding FRB, available in the FRBSTATS Catalogue. Specifically, since the FRB redshift is estimated from the DM, we required that the redshift of the GRB is at least lower than the redshift inferred for the FRB. Only two matches satisfy this criterion:
\begin{itemize}
    \item[a)] \textbf{GRB 110715A}, a long GRB at z$_{\rm GRB}$=0.82, and \textbf{FRB 20171209A}\footnote{\url{https://www.wis-tns.org/object/20171209a}}, a non-repeating FRB discovered by Parkes with an inferred distance of ${\rm z_{FRB}=1.17}$;
    \item[b)] \textbf{GRB 060502B}, a short GRB at an estimated redshift z$_{\rm GRB}$=0.287, and \textbf{FRB 20190309A}\footnote{\url{https://www.chime-frb.ca/catalog/}}, a non-repeating FRB discovered by CHIME with an inferred distance of ${\rm z_{FRB}=0.32}$.
\end{itemize}

Candidate a was reported by \citet{2020ApJ...894L..22W} with a low significance (2.28-2.55 $\sigma$). This is interesting because based on its characteristics, GRB 110715A was associated with a magnetar central engine \citep{2020ApJ...894L..22W}.

Candidate b was also reported before by \citet{2024ApJ...961...45L} with a chance probability of 0.05. The distance between the two events is 13', which is smaller than 1$\sigma$ ($\sigma=21'$). In addition, the two events have a very similar redshift. However, the redshift of the GRB is that of a massive red galaxy that was suggested to be its host galaxy using associative and probabilistic arguments \citep{2007ApJ...654..878B}. This association is debated, especially due to the extremely large offset \citep{2011MNRAS.413.2004C}.

Therefore, we re-analysed the coincidence of the GRB 060502B with the putative host galaxy. We downloaded the candidate field from the Panoramic Survey Telescope and Rapid Response System 1 (Pan-STARRS1) public archive\footnote{\url{https://outerspace.stsci.edu/display/PANSTARRS/}} in the $r$ band. We computed the probability of a chance coincidence (${\rm P_{cc}}$) for the putative host galaxy with respect to the best afterglow position for GRB 06052B. Following the prescriptions from \citet{2022MNRAS.515.4890O}, given the $r$-band host magnitude of $\sim 19.2$ mag (AB, hereafter) and the angular separation from the XRT localisation of $\sim 17.8"$, we found that ${\rm P_{cc}\simeq 0.10}$, which is higher than the previous estimate \citep{2007ApJ...654..878B}. Since the limiting magnitude in the field is $\sim 23$ mag, we cannot exclude other potential fainter hosts, consistent with the best localisation of the GRB. However, \citet{2007ApJ...654..878B} found no better potential associations in the region even with a deeper image. They reported several sources that are marginally consistent with the XRT position, with magnitudes ranging from 24 to 26 mag, and excluded other possible coincident hosts down to magnitude $\sim 26.5$ mag. The presence of an extremely faint host ($> 26.5$ mag) beneath the position of the GRB, due to a higher redshift or because it is a red dwarf galaxy, as was recently discussed for a sample of short GRBs by \citet{2024ApJ...962....5N}, cannot be ruled out.

In addition to the low probability of an association between GRB 060502B and the putative host galaxy at ${\rm z=0.287}$, \citet{2011MNRAS.413.2004C} showed that a $\gtrsim$70 kpc offset would be extremely difficult to attribute to the birth-kick velocity. They proposed a globular cluster nature for the birth site of the compact object binary system, which would guarantee the possibility of observing significantly larger offsets between the birth and merger sites. We also note that according to the short GRB offset sample by \citet{2022ApJ...940...56F}, GRB 060502B would be one of the outliers in the offset distribution, with <1\% of the events found at >70 kpc.

We performed a further search for an association between GRBs and FRBs with the pre-Swift GRBs with redshift, and we found two matches when we only considered the spatial and temporal information:
\begin{itemize}
    \item \textbf{GRB 030226}, a long GRB discovered by HETE-II at z$_{\rm GRB}$=1.98, and \textbf{FRB 20190303C}, a non-repeating FRB discovered by CHIME with an inferred distance of ${\rm z_{FRB}=1.09}$;
    \item \textbf{GRB 051022}, a long GRB discovered by HETE-II at z$_{\rm GRB}$=0.81, and \textbf{FRB 20190608A}, a non-repeating FRB discovered by CHIME with an inferred distance of ${\rm z_{FRB}=0.68}$;
\end{itemize}
However, neither of these matches is confirmed when accounting for the information on the distance, since in both cases, the FRB is closer than the GRB.

\subsection{Chance probability of the association between fast radio bursts and gamma-ray bursts}\label{sect.chance}

Given the size of the samples of GRBs and FRBs considered in Sect. \ref{sec:archivalsearch}, we estimated the probability that a specific number of GRBs and FRBs are associated just by chance. To do this, we proceeded as follows. 

We focused on the FRBs that were discovered by CHIME (516 out of the 633 FRBs that we considered in Sect. ~\ref{sec:archivalsearch}) in order to better control the simulations with a homogeneous sample that represents more than 80\% of the FRBs of our original sample. We generated a synthetic population of 1276 GRBs and another population of 516 FRBs in the sky. We assumed GRBs and FRBs to have an isotropic and homogeneous distribution in space, and for the FRBs, we restricted our simulations to the northern hemisphere (declination between -11 deg and 90 deg) to take the CHIME observable sky into account (see e.g. \citealt{2021ApJS..257...59C}). We assumed that the uncertainty in the localisation for the GRBs is negligible since the mean value of the accuracy in the localisation for the GRBs detected and localised by Swift/XRT is 1.86". We extracted the uncertainty in the localisation for the FRBs from a Gaussian distribution whose mean ($14.9'$) and standard deviation ($6.2'$) were taken from the distribution for the CHIME FRBs with $\sigma_{\rm FRB} \leq 30'$. We then assigned to each source a redshift, randomly extracted from the observed redshift distribution of Swift GRBs\footnote{\url{https://swift.gsfc.nasa.gov/archive/grb_table/}}  and CHIME FRBs \citep{2021ApJS..257...59C}. We also assigned to each GRB a random occurrence time between November 20, 2004 (the starting time of Swift operations), and March 21, 2023. For FRBs, we considered a time interval between July 25, 2018, which is the starting date of the CHIME FRB catalogue, and November 28, 2022, which corresponds to the most recently detected FRB reported in the catalogue. To have enough statistics, we performed 10$^5$ realisations of these two populations.

For the entire population of GRBs, regardless of the information on the distance, we find that the distribution of the number of matches with FRBs has a mean value of 11.4 (median 11) and a standard deviation of 3.4. When we impose that the FRB follows the GRB, the mean value is 9.8 (median 10) with a standard deviation of 3.1. When we further require that the GRB is closer than the FRB, the mean value of the number of matches is 1.6 (median 1), with a standard deviation of 0.9. 

\begin{figure}[h!]
\centering
\includegraphics[width=0.5\textwidth]{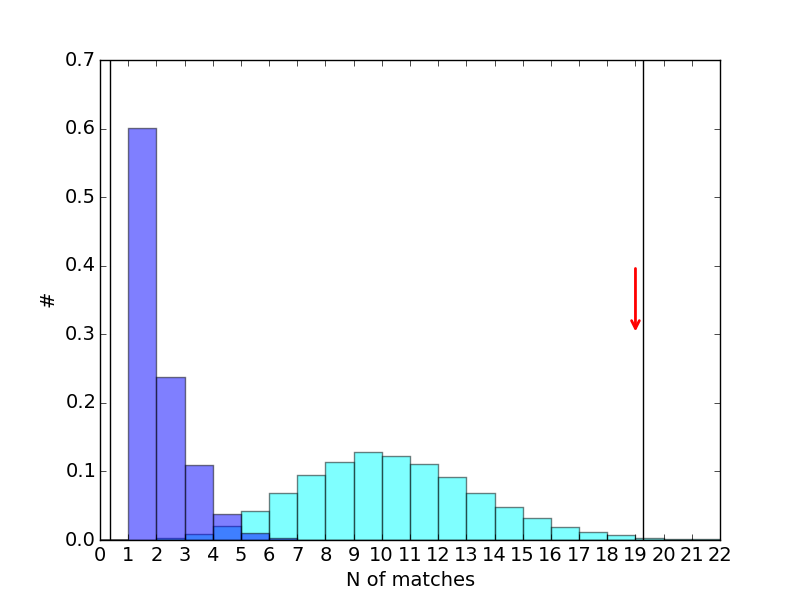}
\caption{Normalised distribution of the number of chance coincidences between GRBs and FRBs obtained by cross-matching the 10$^5$ realisations of GRB and FRB samples when only spatial and temporal constraints (cyan) were applied and when the distance constraint (blue) was also taken into account. The vertical solid lines mark the 3$\sigma$ confidence interval, and the vertical arrow marks   the number of matches using real samples when only spatial and temporal constraints were considered.}\label{Fig:hist}
\end{figure}

In Sect. \ref{sec:archivalsearch} we found only one match between CHIME FRBs and Swift GRBs when we applied the spatial, temporal, and distance constraints. This is consistent with the expected number of chance coincidences. When we only considered spatial and temporal constraints, the number of matches we found (19 CHIME FRBs) is higher than the expected number of chance coincidences (see Fig. \ref{Fig:hist}), although it is still consistent at the 3$\sigma$ level with expectations. 

\section{Can we rule out the association between gamma-ray bursts and fast radio bursts?}\label{sec:limits}

Although there is no clear association between GRBs and FRBs, it may be questioned whether this is sufficient to exclude any connection between them. In other words, we need to determine how likely it is to detect an FRB from a GRB under the hypothesis that all GRBs are associated with FRBs. 

To answer this question, we started from the assumption that every GRB event is associated with an FRB without choosing an a priori time delay because we did not rely on a specific model, and we estimated the detection rate of this population of FRBs. We only considered non-repeating FRBs because they are more likely related to cataclysmic events of the same type as those that stem from GRBs. The inclusion of repeating FRBs requires a detailed modelling of the energy distribution and time-domain behaviour of bursts from a single source, and a dedicated analysis of the rates expected in this scenario will be investigated in a separate work.

To do this, we generated a synthetic population of $10^6$ FRBs and assigned to each of them i) a redshift and ii) a rest-frame isotropic energy. 

i) The redshift was drawn from the redshift distribution of Swift GRBs, and we considered the short and long populations together. This choice was motivated by the fact that we focused on the FRBs that might be associated with GRBs, and these might only represent a subsample of the whole FRB population. 

ii) The rest-frame isotropic energy was  drawn from the FRB energy distribution $\Phi(E)$. The FRB luminosity and energy distributions were investigated in the past by several authors, and they were typically assumed to follow a Schechter function \citep{1976ApJ...203..297S}. For instance, \citet{2020MNRAS.494..665L} used a heterogeneous sample of 46 FRBs (both repeating and non-repeating) that were observed with  different instruments: the Parkes, Arecibo, and Green Bank Telescopes, UTMOST, and the Australian Square Kilometre Array Pathfinder (ASKAP). They measured the FRB luminosity function with a Bayesian approach and took selection effects such as the survey sensitivity into account. Assuming a Schechter  function and neglecting the cosmic evolution of FRBs, they found a slope of the luminosity function $\Phi(L)$ of $\alpha$=-1.79$^{+0.31}_{-0.35}$ and a lower cut-off luminosity $L_0 < 9.1 \times 10^{41}$ erg s$^{-1}$. \citet{2023ApJ...944..105S} used the full sample of 536 FRBs from the first CHIME catalogue \citep{2021ApJS..257...59C}. They  modelled the FRB energy distribution with a Schechter function and assumed that $\Phi(E)$ does not evolve with redshift. They found a differential power-law index of $\alpha=-1.3^{+0.7}_{-0.4}$ and a characteristic exponential cut-off energy of $E_{char}=2.38^{+5.35}_{-1.64} \times 10^{41}$ erg. \citet[][hereafter H22]{2022MNRAS.511.1961H} used a  homogeneous subsample of FRBs from the first CHIME catalogue. They divided the selected FRBs into repeaters and non-repeaters and then into several subsamples, filling different redshift bins that covered the range between 0.05 and 3.6. Because the results may depend on how the redshift bins are selected, they performed their analysis considering two different sets of redshift bins (called redshift A and redshift B) to take this uncertainty into account. They fitted  Schechter functions to the derived energy functions and took the redshift evolution into account. They found $\alpha$=-1.4$^{+0.7}_{-0.5}$ (-1.1$^{+0.6}_{-0.4}$) for redshift bin A (redshift bin B). Similar results were also presented in other papers (see e.g. \citealp{2019ApJ...883...40L}). Despite the different methods and different FRB samples used, the reported results broadly agree within the errors. It is important to note that the observed energy and luminosity distributions were allowed to include the contribution from different FRB populations (e.g. with different progenitors), which might have made individual distributions different from each other.

In this work, we used the energy distribution derived by H22, and, as already mentioned, we considered non-repeating  FRBs.  
Since the redshift range of GRBs is wider than the range considered in H22, we assumed that the energy distribution  of FRBs with z < 0.05 (> 3.6) was the same as obtained for the lowest (highest) redshift bin, and we considered redshift A and redshift B cases, with the parameters reported in Table 1 of H22. 
We focused on the CHIME survey because it currently has the best combination in terms of sensitivity and field of view (fov).  We then associated a rest-frame isotropic energy integrated over the bandwidth $\Delta\nu=$400 MHz (the CHIME frequency width), $E_{\rm{rest,400}}$, with each FRB. We assumed a fiducial value for the minimum FRB rest-frame energy of 10$^{37}$ ergs, which is the minimum energy of the FRBs in the CHIME catalogue, and a maximum value of 10$^{50}$ ergs. 

From these values, we computed  the observed fluence $F_{\nu}$ in the CHIME frequency band (400 MHz - 800 MHz) as 
\begin{equation}
{\rm F_\nu=\frac{(1+z)^{2-\gamma} E_{\rm{rest,400}}}{4 \pi d_L^2(z) \Delta\nu},}
\end{equation}
where $d_L(z)$ is the luminosity distance at redshift z, estimated with the cosmological parameters from \citet{2016A&A...594A..13P}, $\Delta\nu$ is the frequency bandwidth (taken as 400 MHz), and $\gamma$ is the FRB spectral index (F $\propto \nu^{-\gamma}$), which we assumed to be equal to 1.4 (see e.g. \citealp{2023ApJ...944..105S}).


\begin{figure*}
\centering
\includegraphics[width=0.45\textwidth]{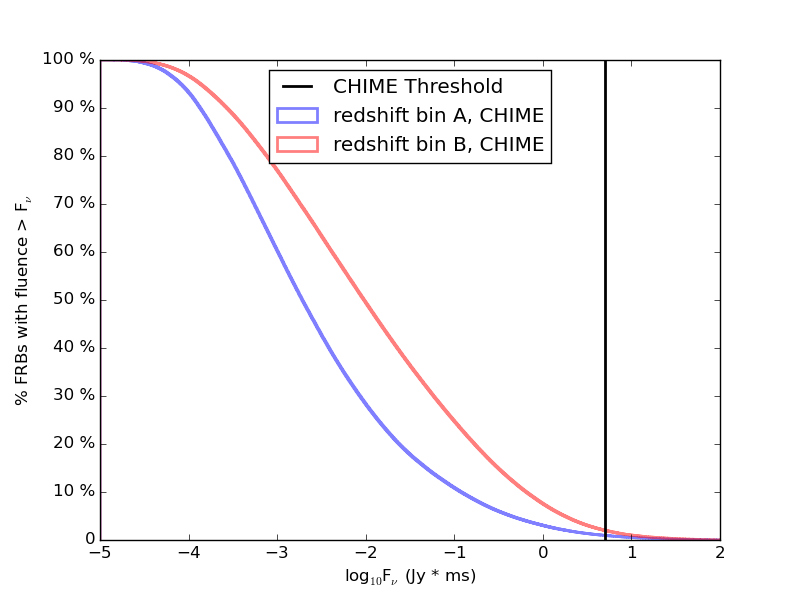} \includegraphics[width=0.45\textwidth]{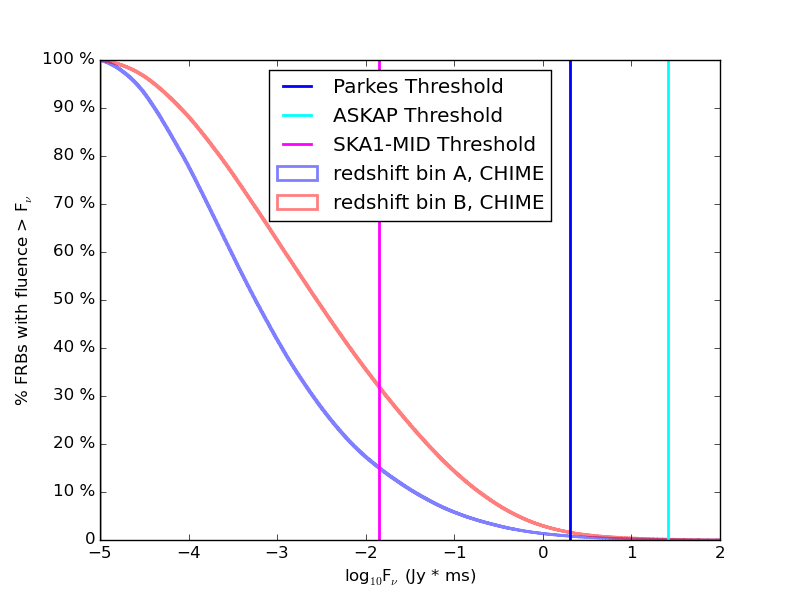}
\caption{Percentage of simulated FRBs with a fluence greater than or equal to F$_{\nu}$ vs. F$_{\nu}$  for redshift bin A (in blue) and redshift bin B (in red). F$_{\nu}$ has been computed in the CHIME frequency band (left) and at 1.4 GHz (right). The vertical coloured lines mark the position of the assumed fluence threshold for CHIME, Parkes, ASKAP, and SKA1-MID. }\label{fig:NvsFnu}
\end{figure*}

Then, to evaluate how many FRBs caused by GRBs can be detected with CHIME, we compared the values of $F_\nu$ with the completeness threshold (95$\%$ c.l.) of ${\rm F_{lim}=5}$ Jy ms to account for the different sources of the sensitivity variation \citep{2021ApJS..257...59C}. 
The percentage of simulated FRBs with a fluence greater than ${\rm F_{lim}=5}$ Jy ms is 1\% for redshift bin A and 2\% for redshift bin B (see Fig. \ref{fig:NvsFnu}, left panel).

We then estimated the detection rate of FRBs associated with GRBs detected by Swift (${\rm R_{FRB}}$) by multiplying these percentages by the Swift average detection rate (85 GRBs yr$^{-1}$ discovered by BAT and 70 GRBs yr$^{-1}$ detected also by XRT)\footnote{\url{https://swift.gsfc.nasa.gov/archive/grb_table/}} and an instantaneous fov of $120^\circ\times 2^\circ$ \citep{2021ApJS..257...59C}, obtaining ${\rm R_{FRB}= [5-11]\times10^{-3}\,yr^{-1}}$, where the lower and upper boundaries correspond to redshift bin A and B, respectively (see Table~\ref{tab:par}). 
These detection rates were obtained by assuming that each GRB is associated with one FRB, and this implicitly takes into account that there is a model-dependent time delay between the two events. 
Despite many uncertainties in the modelling of the FRB population and the simplified description of the detection process used, this result shows that the absence of a clear association between FRBs in the current (4 years) CHIME catalogue and Swift GRBs cannot exclude that the two phenomena can have a common progenitor.

We then performed the same analysis considering Parkes and ASKAP. In this case, we evaluated the detection rates by calculating the observed fluence $\rm{F_\nu}$ at 1.4 GHz and comparing it to the completeness threshold (95$\%$ c.l.) of ${\rm F_{lim}=}$ 2 and 26 Jy ms, respectively \citep{2018MNRAS.475.1427B,2018Natur.562..386S}. The percentage of simulated FRBs with a fluence greater than ${\rm F_{lim}}$ is 0.8\% and 0.1\% for redshift bin A, and it is  1.7\% and 0.3\% for redshift bin B (see Fig. \ref{fig:NvsFnu}, right panel). 

For Parkes, we considered a duty cycle of 100\% and a fov of 0.6 deg$^2$ \citep{2020MNRAS.494..665L} and obtained ${\rm R_{FRB}}$= [1-2]$\times10^{-5}$\,yr$^{-1}$, where the lower and upper boundaries correspond to redshift bin A and B, respectively. For ASKAP, we considered a duty cycle of 100\% and a fov of 150 deg$^2$ \citep{2020MNRAS.494..665L} and obtained ${\rm R_{FRB}= [4-8]\times10^{-4}\,yr^{-1}}$, where the lower and upper boundaries correspond to redshift bin A and B, respectively (see table ~\ref{tab:par}).

\begin{table}[h!]
\caption{\label{t7}Parameters used to estimate the FRB detection rates.}
\begin{center}
\begin{tabular}{c|c|c|c|c|c} 
\hline\hline
& ${\rm F_{lim}}$ & DC  & fov & ${\rm R_{FRB}}$ & Refs. \\
& Jy ms & & deg$^2$ & yr$^{-1}$ & \\
\hline
CHIME & 5 & 100 & 240 & [5-11]$\times10^{-3}$ & 1\\
\hline\hline
Parkes & 2 & 100 & 0.6  & [1-2]$\times10^{-5}$ & 2\\ 
ASKAP & 26 & 100 & 150 & [4-8]$\times10^{-4}$ & 3\\ 
\hline\hline
SKA1-MID & 0.014 & 20 & 20 & [1-3]$\times10^{-3}$ & 4\\ 
\end{tabular}
\tablefoot{Parameters include the detection threshold (${\rm F_{lim}}$), the duty cycle (DC), and the fov of  CHIME, Parkes, ASKAP, and SKA1-MID.}
\tablebib{(1) \cite{2021ApJS..257...59C,2020MNRAS.494..665L}; (2) \cite{2018MNRAS.475.1427B,2020MNRAS.494..665L}; (3) \cite{2018Natur.562..386S,2020MNRAS.494..665L}; (4) \cite{2015aska.confE..51F}.}
\end{center}
\label{tab:par} 
\end{table}


It is important to highlight that  the rates obtained in this work might underestimate the actual rates because we did not consider the population of repeating FRBs in our analysis. The inclusion of repeating FRBs could increase the detection rates because we would have multiple FRBs for each GRB.

\section{Discussion and conclusions} \label{sec:conclusions}

We performed a comprehensive search for a possible association between GRBs and FRBs by cross-matching Swift/GRBs with all the FRBs reported in the FRBSTATS catalogue. We initially applied only spatial and temporal constraints association, and then we also considered the distance information of the subsample  of Swift/GRBs with a known redshift. In this last case, we identified two matches: a) GRB 110715A/FRB 20171209A, and b) GRB 060502B/FRB 20190309A. In previous searches \citep{2020ApJ...894L..22W}, candidate a was reported as a low-significance match. Candidate b was also reported by \cite{2024ApJ...961...45L} with a statistical significance $< 3 \sigma$. In addition the redshift estimate for the GRB is debated, which further weakens the association. In any case, the number of matches found in our searches is consistent at the 3$\sigma$ level with the expectations from chance coincidences. 

The lacking unambiguous cross-match between the GRB and FRB catalogues does not mean that the two populations are not connected. Even under the hypothesis that all GRBs are associated with non-repeating FRBs and that their collimation is such that it does not prevent a coincident detection (e.g. \citealt{2021MNRAS.501.3184S}), a survey with the same characteristics as CHIME would take hundreds of years to detect at least one FRB associated with a GRB that was discovered by Swift, as shown in Sect. \ref{sec:limits}. The expectations are even less promising for other current facilities such as Parkes and ASKAP.   

Our results are valid for any delay time between GRBs and FRBs. 
However, our analysis can only provide constraints on models that predict a time delay between the two events from zero to a few dozen years, assuming a typical expected lifetime of the instruments considered in this work. Much longer time delays cannot be probed with direct observations of the two phenomena.

The Square Kilometre Array (SKA) is a planned large radio interferometer that is designed to operate over a wide range of frequencies, and its sensitivity and survey speed are an order of magnitude greater than those of any current radio telescope. The SKA will comprise two separate arrays, one in Western Australia and the other in South Africa, and it is designed to be built in phases. The first (SKA1) is expected to become fully operational by the late 2020s and consists of three elements: the first  element operating at low frequency (SKA1-LOW, [50-350] MHz),
the second element at intermediate/high frequency (SKA1-MID, [1.2-1.7] GHz), and the third element will be optimized for surveys (SKA1-survey). For SKA1-MID, considering a completeness threshold of ${\rm F_{lim}=0.014}$ Jy ms \citep{2015aska.confE..51F}, we obtain ${\rm R_{FRB}= [1-3]\times10^{-3}\,yr^{-1}}$, where the lower and upper boundaries correspond to redshift bin A and B, respectively (see table~\ref{tab:par}). Therefore, despite its much higher sensitivity, the expectations for a joint detection are comparable to CHIME performances because the fov is smaller. However, we here considered the population of CHIME FRBs as representative of the whole population(s) of FRBs, which is not necessarily the case. With its higher sensitivity, SKA will probe the faint end of the FRB energy distribution and might discover new features that are not accounted for in our model.

The rates were also derived assuming that GRBs are associated with non-repeating FRBs. If instead they were connected with the population of repeaters, we would have multiple FRBs for one GRB, which would accordingly increase the detection probability. As already mentioned in Sect. \ref{sec:limits}, a dedicated analysis of the rates expected in this scenario will be investigated in a separate work.

Another way to increase the probability of having a joint FRB-GRB detection is a more efficient GRB discovery machine, such as the Transient High Energy Sky and Early Universe Surveyor (THESEUS; \citealp{2018AdSpR..62..191A}). The space mission concept THESEUS was selected by ESA as a  candidate mission (launch in 2037) with the aim of exploiting GRBs to study the early Universe and of providing a fundamental contribution to time-domain and multi-messenger astrophysics. THESEUS will detect a factor $\sim$ 10 more GRBs than Swift in the redshift range up to $\sim 2.5$ with an accurate localisation in the sky \citep{2021ExA....52..277G}, which opens the possibility of observing a GRB associated with a detectable FRB by a facility with the same characteristics as CHIME within a lifetime of $\sim$10 yr.

\begin{acknowledgements}
We acknowledge use of the CHIME/FRB Public Database, provided at https://www.chime-frb.ca/ by the CHIME/FRB Collaboration. We thank Paolo D'Avanzo, Sara Elisa Motta and Sergio Campana for useful discussions. 
\end{acknowledgements}

%
%
\bibliographystyle{aa}
\bibliography{bibliography}

\begin{appendix} 

\section{GRBs and FRBs matches}
\begin{table*}[h!]
\caption{\label{t7}List of GRBs and FRBs for which a match has been found.}
\label{tab:list}
\centering   
\begin{tabular}{c c c c c | c c c c c c |c} 
\hline\hline 
GRB & z$_{\rm GRB}$ & RA & Dec & err & FRB & z$_{\rm FRB}$ & DM & RA & Dec & err  & $\verb|#|  \sigma$\\
 & & deg & deg & " & & & & deg & deg & ' & \\
\hline    
160705B &        - &  168.10942 &   46.69989 &        1.5 & 20180907E &        0.4 &      381.7 &     167.88 &      47.09 &      19.94 & 1.3\\
151006A &        - &  147.42558 &   70.50303 &        1.5 & 20190612A &        0.4 &      427.0 &     148.16 &      70.42 &      17.40 & 0.9\\
150314A &        - &  126.67042 &   63.83431 &        1.7 & 20190409B &        0.3 &      297.6 &     126.65 &      63.47 &      18.25 & 1.2\\
140713A &        - &  281.10592 &   59.63347 &        1.4 & 20190429A &        0.4 &      470.7 &     281.09 &      59.42 &      20.36 & 0.6\\
140211A &        - &  124.22329 &   20.24319 &        4.4 & 20190224C &        0.4 &      495.0 &     124.05 &      19.78 &      17.84 & 1.7\\
120213A &        - &  301.01212 &   65.41131 &        1.4 & 20180918A &        1.3 &     1454.1 &     301.27 &      64.96 &      17.84 & 1.6\\
110801A &        1.9 &   89.43596 &   80.95631 &        1.5 & 20190208B &        0.7 &      713.3 &      91.00 &      80.88 &       8.08 & 1.9\\
110223A &        - &  345.85217 &   87.55786 &        2.0 & 20190519H &        1.1 &     1169.5 &     342.99 &      87.37 &      11.29 & 1.2\\
110223A &        - &  345.85217 &   87.55786 &        2.0 & 20190609A &        0.3 &      315.4 &     345.30 &      87.94 &      25.97 & 0.9\\
100526A &        - &  230.76904 &   25.63219 &        1.7 & 20181221A &        0.3 &      313.8 &     230.58 &      25.86 &      17.40 & 1.0\\
100316A &        - &  251.97875 &   71.82708 &        2.2 & 20190409C &        0.6 &      674.5 &     252.60 &      71.62 &      19.53 & 0.9\\
090904A &        - &  100.88458 &   50.20289 &        1.6 & 20190317B &        0.4 &      425.4 &     101.01 &      49.73 &      12.71 & 2.3\\
090813 &        - &  225.78783 &   88.56822 &        1.4 & 20190425B &        1.0 &     1030.3 &     210.12 &      88.60 &      13.96 & 1.7\\
090813 &        - &  225.78783 &   88.56822 &        1.4 & 20190609B &        0.3 &      292.8 &     210.49 &      88.35 &       9.86 & 2.8\\
070429B &        0.9 &  328.01587 &  -38.82833 &        2.4 & 20180130A$^a$ &        0.3 &      343.5 &     328.05 &     -38.57 &      17.20 & 0.9\\
060912A &        - &    5.28387 &   20.97181 &        1.4 & 20190316A &        0.5 &      516.0 &       5.23 &      20.51 &      19.53 & 1.4\\
060515 &        - &  127.28892 &   73.56778 &        3.5 & 20181213A &        0.6 &      677.7 &     127.66 &      73.87 &      13.42 & 1.4\\
060510B &        4.9 &  239.12192 &   78.56989 &        1.5 & 20181017B &        0.3 &      304.1 &     237.76 &      78.50 &      20.38 & 0.8\\
060319 &        - &  176.38767 &   60.01081 &        1.4 & 20181214C &        0.6 &      632.4 &     175.93 &      60.02 &       7.75 & 1.8\\
\hline     
110715A &        0.8 & 237.68358 & -46.23531 &      1.4 &  20171209A$^b$ &        1.2 &     1457.4 &     237.60 &     -46.17 &      10.61 &  0.5\\
060502B &        0.3 & 278.93846 &  52.63153 &      5.2 &  20190309A &        0.3 &      357.5 &     278.96 &      52.41 &      19.94 & 0.7\\
\hline
\end{tabular}
\tablefoot{The list contains the 21 positive matches found considering spatial and temporal constraints. The ones listed in the lower panel are those that also satisfy the distance constrain. All FRBs have been detected by CHIME with the exception of $^a$ from ASKAP and $^b$ from Parkes. The last column report the distance over $\sigma$ between FRBs and GRBs.}
\end{table*}

\end{appendix}

\end{document}